\documentclass[twocolumn,setspace,prc,showpacs,amsmath,amssymb,superscriptaddress,letterpaper]{revtex4-1}
\usepackage{hyperref}
\hypersetup{
     colorlinks,
     linkcolor={red!50!black},
     citecolor={blue!50!black},
     urlcolor={blue!80!black}
}
\usepackage[table]{xcolor}
\usepackage{epsfig}
\usepackage{dcolumn}
\usepackage{bm}
\begin{document}

\title{A proton density bubble in the doubly-magic $^{34}$Si nucleus}
\author{A.~Mutschler}
\affiliation{Institut de Physique Nucl\'eaire, IN2P3-CNRS,
F-91406 Orsay Cedex, France}
\affiliation{Grand Acc\'el\'erateur National d'Ions Lourds (GANIL),
CEA/DSM - CNRS/IN2P3, B.\ P.\ 55027, F-14076 Caen Cedex 5, France}

\author{A.~ Lemasson}
\affiliation{Grand Acc\'el\'erateur National d'Ions Lourds (GANIL),
CEA/DSM - CNRS/IN2P3, B.\ P.\ 55027, F-14076 Caen Cedex 5, France}
\affiliation{National Superconducting Cyclotron Laboratory, Michigan State University, East Lansing, Michigan 48824, USA}

\author{O.~Sorlin}
\affiliation{Grand Acc\'el\'erateur National d'Ions Lourds (GANIL),
CEA/DSM - CNRS/IN2P3, B.\ P.\ 55027, F-14076 Caen Cedex 5, France}

\author{D. Bazin}
\affiliation{Department of Physics and Astronomy and National Superconducting
Cyclotron Laboratory, Michigan State University, East Lansing, Michigan, 48824-1321, USA}

\author{C.\ Borcea}
\affiliation{IFIN-HH, P. O. Box MG-6, 76900 Bucharest-Magurele, Romania}

\author{R.\ Borcea}
\affiliation{IFIN-HH, P. O. Box MG-6, 76900 Bucharest-Magurele, Romania}

\author{Z. Dombr\'adi}
 \affiliation{Institute for Nuclear Research, Hungarian Academy of Sciences, P.O. Box 51, Debrecen, H-4001, Hungary}

\author{J.-P. Ebran}
\affiliation{CEA, DAM, DIF, F-91297 Arpajon, France}

\author{A. Gade}
\affiliation{Department of Physics and Astronomy and National Superconducting
Cyclotron Laboratory, Michigan State University, East Lansing, Michigan, 48824-1321, USA}


\author{H. Iwasaki}
\affiliation{Department of Physics and Astronomy and National Superconducting
Cyclotron Laboratory, Michigan State University, East Lansing, Michigan, 48824-1321, USA}

\author{E. Khan}
\affiliation{Institut de Physique Nucl\'eaire, IN2P3-CNRS,
F-91406 Orsay Cedex, France}

\author{A.~Lepailleur}
\affiliation{Grand Acc\'el\'erateur National d'Ions Lourds (GANIL),
CEA/DSM - CNRS/IN2P3, B.\ P.\ 55027, F-14076 Caen Cedex 5, France}

\author{F. Recchia}
\affiliation{National Superconducting Cyclotron Laboratory, Michigan State
  University, East Lansing, Michigan 48824, USA}

\author{T. Roger}
\affiliation{Grand Acc\'el\'erateur National d'Ions Lourds (GANIL),
CEA/DSM - CNRS/IN2P3, B.\ P.\ 55027, F-14076 Caen Cedex 5, France}

\author{F.~ Rotaru}
\affiliation {IFIN-HH, P. O. Box MG-6, 76900
Bucharest-Magurele, Romania}

\author{D. Sohler}
 \affiliation{Institute for Nuclear Research, Hungarian Academy of Sciences, P.O. Box 51, Debrecen, H-4001, Hungary}

\author{M.\ Stanoiu}
\affiliation {IFIN-HH, P. O. Box MG-6, 76900
Bucharest-Magurele, Romania}

\author{S. R. Stroberg}
\affiliation{Department of Physics and Astronomy and National
Superconducting Cyclotron Laboratory, Michigan State University,
East Lansing, Michigan, 48824-1321, USA}
\affiliation{TRIUMF, 4004 Westbrook Mall, Vancouver, British Columbia, V67 2A3 Canada}

\author{J.\ A.\ Tostevin}
     \affiliation{Department of Physics, University of Surrey,
       Guildford, Surrey GU2 7XH, United Kingdom}

\author{M. Vandebrouck}
\affiliation{Institut de Physique Nucl\'eaire, IN2P3-CNRS,
F-91406 Orsay Cedex, France}

\author{D. Weisshaar}
\affiliation{National Superconducting Cyclotron Laboratory, Michigan State University, East Lansing, Michigan 48824, USA}

\author{K.~Wimmer}
\affiliation{National Superconducting Cyclotron Laboratory, Michigan State University, East Lansing, Michigan 48824, USA}
\affiliation{Department of Physics, Central Michigan University, Mt. Pleasant, Michigan 48859, USA}
\affiliation{Department of Physics, The University of Tokyo, Hongo, Bunkyo-ku, Tokyo 113-0033, Japan}

\pacs{to be defined}
\date{\today}
\maketitle

{ \bf{Many properties of the atomic nucleus, such as vibrations, rotations and
incompressibility can be interpreted as due to a two-component quantum
liquid of protons and neutrons. Electron scattering measurements
on stable nuclei demonstrate that their central densities are saturated, as for liquid drops. In exotic nuclei near the limits of
mass and charge, with large imbalances in their proton and neutron numbers,
the possibility of a depleted central density, or a ``bubble" structure,
was discussed in a recurrent manner since the seventies.
Here we report first experimental evidence that points to a depletion of the central density
of protons in the short-lived nucleus $^{34}$Si. The proton-to-neutron density
asymmetry in $^{34}$Si offers the possibility to place constraints on the density
and isospin dependence of the spin-orbit force - on which nuclear models
have disagreed for decades- and on its stabilizing effect towards limits of nuclear existence.}}

Microscopic systems composed of atoms or clusters can exhibit intrinsic
structures that are bubble-like, with small or depleted central densities.
For example, the fullerene molecules, composed of C atoms, are structures
with extreme central depletion~\cite{fullerene}. In nuclear physics,
depletions also arise in nuclei with well-developed cluster structures
when clusters are arranged in a triangle or ring-like structure - such as
in the triple-$\alpha$ Hoyle state~\cite{hoyle,Freer}. Unlike such a
non-homogeneous, clustered system, central density depletions or bubble-like
structures would be much more surprising in homogeneous systems, such as
typical atomic nuclei with properties characteristic of a quantum liquid
~\cite{ebran}.

This hindrance of bubble formation in atomic nuclei is inherent in the nature
of the strong force between nucleons, which is strongly repulsive at short
distances (below 0.7 fm), attractive at medium range ($\approx$1.0 fm) and
vanishes at distances beyond 2 fm. 
In a classical picture, the medium-ranged attraction of nuclear forces implies that nucleons interact strongly and attractively only with
immediate neighbors, leading to a saturation of the nuclear central density,
$\rho_0$.  Quantum mechanically, the delocalization of nucleons~\cite{mott}
leads to a further homogeneity of the density. Extensive precision electron
scattering studies from stable nuclei~\cite{scattering} confirm that their
central densities are essentially constant, with $\rho_0\approx 0.16$ fm$^{-3}$,
independent of the number of nucleons $A$. As a consequence, like a liquid
drop, the nuclear radii and volumes increase as $A^{1/3}$ and as $A$, respectively. Thus,
{\it a priori}, bubble-like nuclei with depleted central densities are
unexpected.

Historically, the possibility of forming bubble nuclei was investigated
theoretically in intermediate-mass ~\cite{campi,Davies73,khan08,Grasso},
superheavy ~\cite{Bender} and hyperheavy systems~\cite{Decharge}. In general,
central depletions will arise from a reduced occupation of single particle
orbits with low angular momentum $\ell$. These wave functions extend throughout the nuclear
interior whereas those with high-$\ell$ are more excluded by centrifugal
forces. For example, in a comparison of the charge densities of $^{206}$Pb
and $^{205}$Tl, the contribution from $\ell$=0 orbits (there $3s$)
is peaked at the nuclear center \cite{Cavedon}. The amplitude of this
central depletion in $^{205}$Tl is of order 11\%. A much larger central
depletion of protons, of about 40\% compared to stable $^{36}$S, was
proposed in $^{34}$Si \cite{Grasso,Li} using various mean field approaches,
arising from the proton occupancy of the $2s_{1/2}$ orbital. However, recent theoretical calculations 
suggest that nuclear correlations act to smoothen these orbital occupancies in both
the heavy and superheavy nuclei \cite{Wang15,Afa} and in $^{34}$Si \cite{Yao}. Here, we use the one-proton
removal $(-1p)$ reaction technique to show that the $2s_{1/2}$ proton
orbit in $^{34}$Si is in fact essentially empty, in contrast to $^{36}$S
where this $2s_{1/2}$ orbit is almost fully occupied by 1.7(4) protons compared to the maximum occupancy of 2 \cite{Khan85,36S}.


A beam of $4 \times 10^5$ $^{34}$Si nuclei per second was produced by the
fragmentation of a 140 MeV/u $^{48}$Ca primary beam on a 846 mg$\cdot$cm$
^{-2}$ thick $^9$Be target at the Coupled Cyclotron Facility at the National
Superconducting Cyclotron Laboratory. The $^{34}$Si then impinged on a
$^9$Be secondary target (100 mg$\cdot$cm$^{-2}$) producing $^{33}$Al nuclei
through the $(-1p)$ reaction. These $^{33}$Al residues were identified through
their measured energy-loss in an ionization chamber located at the focal plane
of the S800 spectrograph, and their time-of-flight between two scintillators
placed at the object and image focal planes of the device. Their trajectories
were obtained from their positions measured at two cathode-readout drift
chambers.

Prompt $\gamma$-rays, originating from the in-flight decay of excited $^{33}$Al
produced during the reaction, were detected in coincidence with the $^{33}$Al
residues in the seven modules of the GRETINA array \cite{Paschalis13} that
surrounded the target at angles near 90$^\circ$ and 58$^\circ$. Event-by-event
Doppler reconstruction was performed using the deduced $^{33}$Al velocity at
the mid-target position, the position reconstruction on the target, and the
$\gamma$-ray detection angle -- determined from the position of the greatest
energy deposition in the GRETINA array. An absolute in-flight efficiency of 6.5\%
and an energy resolution of $\sigma\simeq 2$ keV, respectively, were obtained at
1 MeV, based on the use of calibrated sources and GEANT4 simulations
\cite{Agostinelli03} to account for the Lorentz boost. A systematic uncertainty of 0.25\% is estimated
on the $\gamma$ energy centroid.

\begin{figure} [h]
\includegraphics[width=\columnwidth] {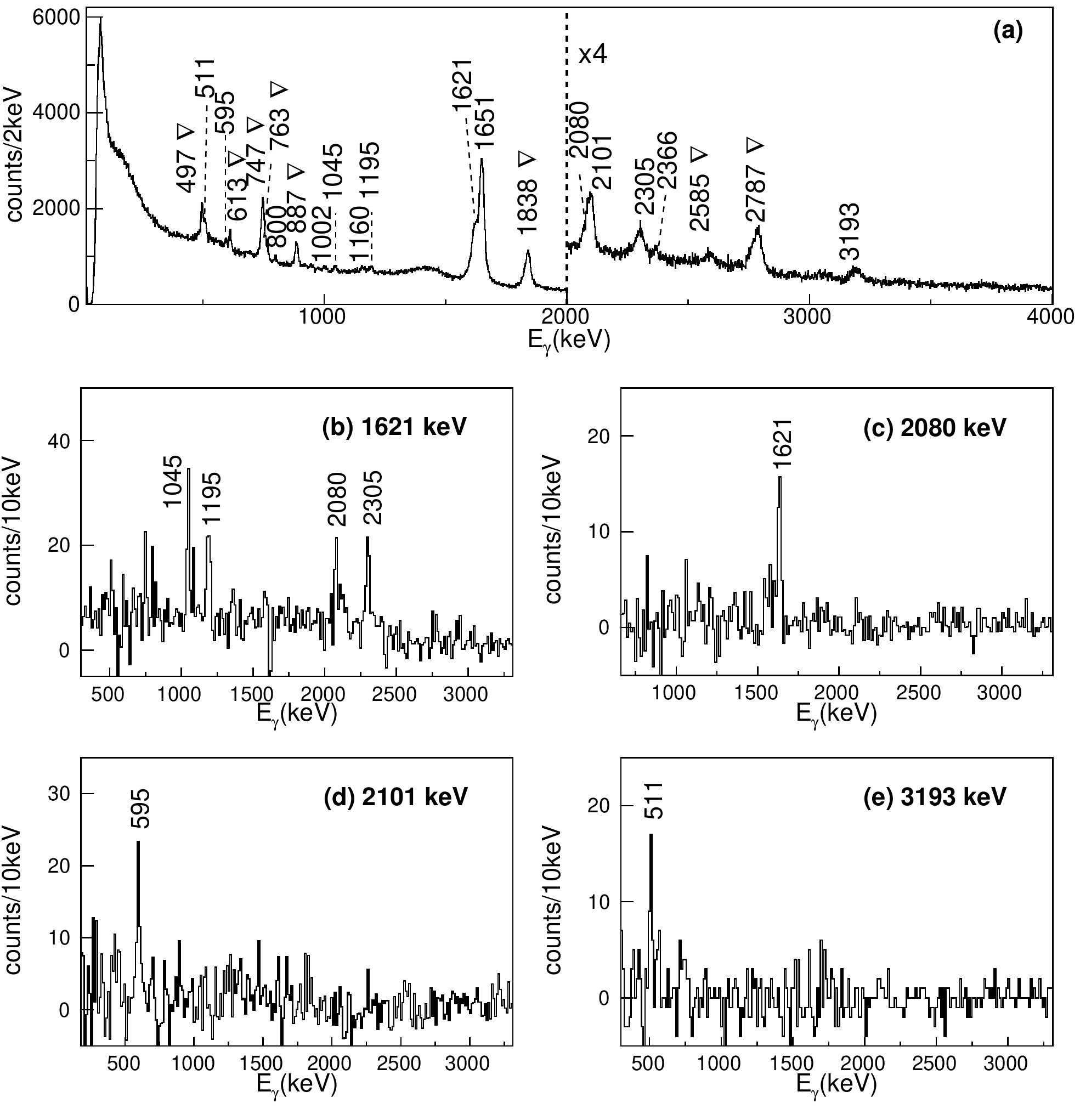}
\caption {{\bf Gamma-ray spectra of $^{33}$Al.} Singles (first row) and $\gamma$-gated (second to third rows)
Doppler-corrected $\gamma$-ray spectra of $^{33}$Al. The $\gamma$-rays
labeled with an empty triangle, accounting for 5.8 \% of the cross section,
are not produced by the one-proton removal mechanism and are not relevant
for determining the proton occupancies. Their origin is discussed in Ref.
\cite{36S}. } \label{33Al-gamma}
\end{figure}

\begin{figure*}
\begin{minipage} {9cm}
\includegraphics[height=7cm]{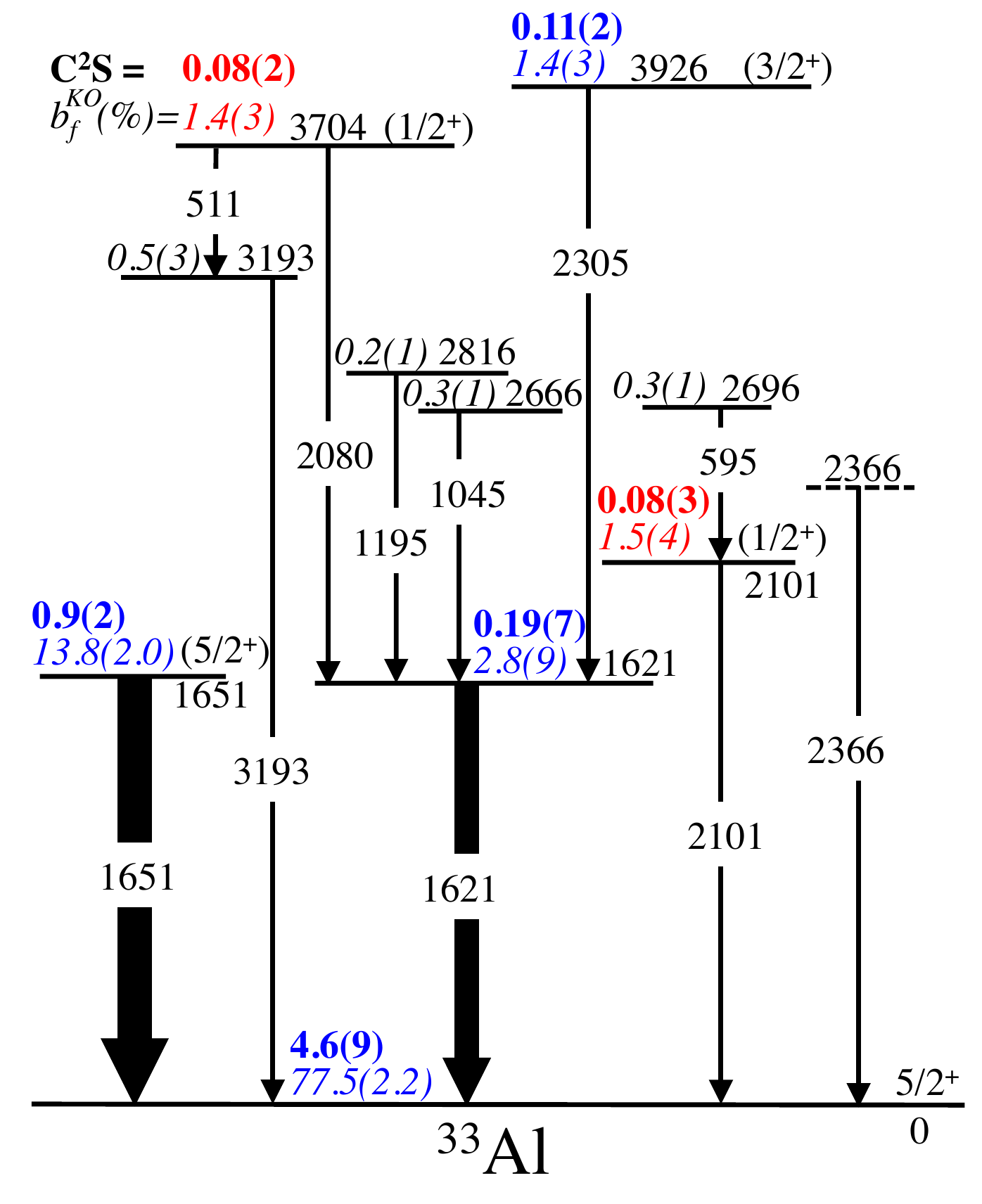}
\end{minipage}
\begin{minipage} {7cm}
\vspace{0.8cm}
\includegraphics[height=7cm,width=7cm]{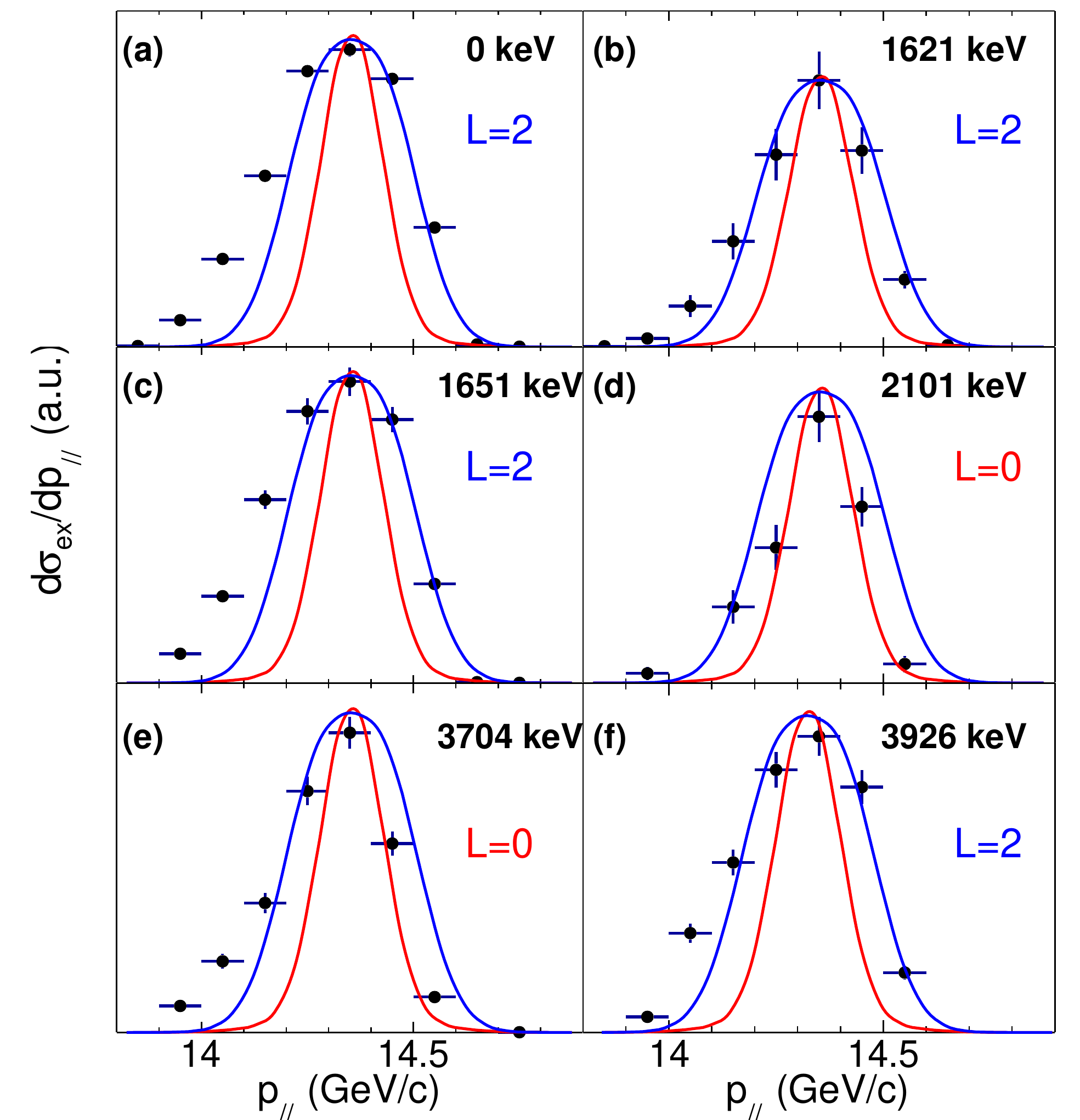}
\end{minipage}
\caption{ {\bf Level scheme of $^{33}$Al with parallel momentum distributions of the strongest populated states.}
Left: Level scheme of $^{33}$Al, obtained from the $\gamma$-$\gamma$
coincidence spectra of Fig. \ref{33Al-gamma}, with energies (in keV) and branching
ratios $b_f^{KO}$ (in $\%$). Error bars on the $b_f^{KO}$ values result from the uncertainty in extracting the intensity of the $\gamma$ transitions decaying from the corresponding levels. When these levels are fed from higher lying states, the corresponding feeding contribution was subtracted, inducing larger error bars. As fed by many transitions, the branching ratio to the ground state therefore has the largest error bar.  The J$^\pi$ assignments and experimental spectroscopic
factors $C^2 S^{exp}_{norm}$ of the strongest populated states are
shown. Uncertainties on the $C^2 S^{exp}_{norm}$ values are derived from Equation \ref{eq1}. They include those on $b_f^{KO}$ discussed above and on the empirical quenching factor $R_S$, that amounts to about 20\%.
Right (a-f): Experimental parallel momentum, $p_{\parallel}$, distributions for the
strongest populated states in $^{33}$Al (black crosses) are compared to theory,
assuming removal of an $\ell$=0 (red curves) or $\ell$=2 (blue curves) proton
from the $^{34}$Si ground state. As explained in Section 'Methods', the
high momentum part of $p_{\parallel}$ is considered in this comparison. Momentum
distributions for weakly populated states ($b_f^{KO} <1$\%) have insufficient
statistics to be exploited. Horizontal bars correspond to the binning on the $p_{\parallel}$ value. Vertical error bars are deduced from uncertainties on $b_f^{KO}$  per bin of $p_{\parallel}$ value. }
\label{33Al-ls}
\end{figure*}

The $\gamma$ singles (first row of Fig. \ref{33Al-gamma}), $\gamma - \gamma$
coincidences (second to third rows of Fig. \ref{33Al-gamma}), and the relative
$\gamma$ intensities were used to establish the level scheme of excited states in $^{33}$Al,
shown in the left part of Fig. \ref{33Al-ls}. Energies and branching ratios
are given for each populated $^{33}$Al final state. The
energy of seven $\gamma$-rays match, within uncertainties, those observed
in the $\beta$-decay of $^{33}$Mg \cite{Tripa08}. However, the level scheme
proposed in Fig. \ref{33Al-ls} differs significantly from that of Ref.
\cite{Tripa08} where, unlike in the present work, $\gamma -\gamma$ coincidences
were rarely exploited.

The orbital angular momenta $\ell$ of the protons removed from $^{34}$Si
are determined by comparing, in the right panel of Fig. \ref{33Al-ls},
the experimental ($\gamma$-gated) and theoretical longitudinal momentum
distributions ($p_{\parallel}$) of the $^{33}$Al. The latter are described in
the Section 'Methods'. The ground state momentum distribution in Fig.
\ref{33Al-ls} is obtained by subtracting contributions from excited
states. The $p_{\parallel}$ distributions of the 0-, 1621-, and 1651-keV states
are characteristic of $\ell$=2 proton removal. The much narrower $p_{\parallel}$
of the 210- and 3704-keV states suggest $\ell$=0 assignments.

\begin{figure}[h]
\begin{center}
\includegraphics[width=8.5cm]{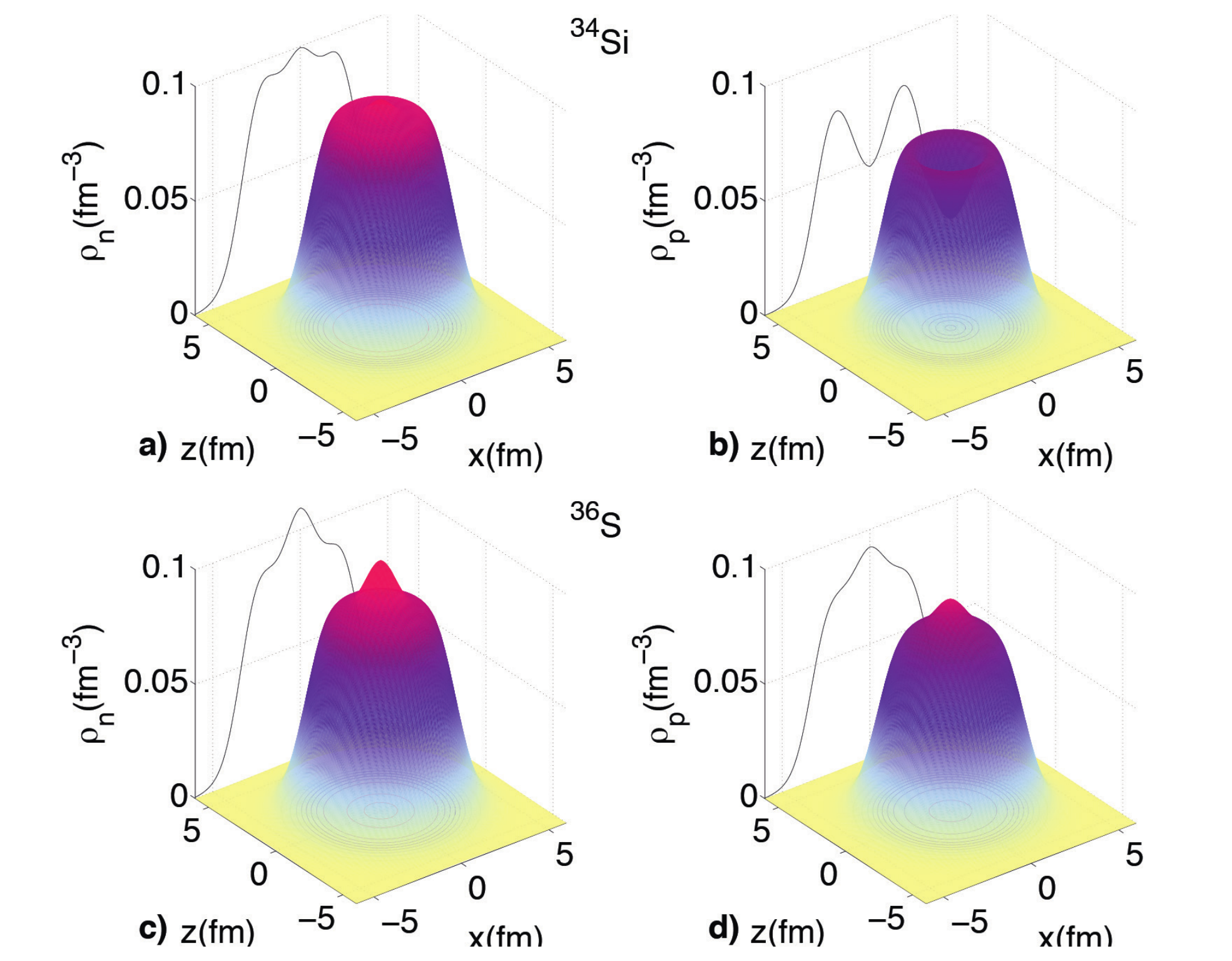}
\caption{{\bf Neutron and proton density distributions of the $^{34}$Si and $^{36}$S nuclei.} Neutron (a) and proton (b) density distributions of $^{34}$Si computed using
RHFB calculations with the PKO2 interaction. (c) and (d) are those for
$^{36}$S. While the proton and neutron density distributions are similar
in $^{36}$S, they are significantly different in $^{34}$Si, with a
sizeable central depletion of protons. Orbital occupancies obtained from
these calculations are very similar to those deduced experimentally,
providing a visualization of the proton and neutron density distributions in the two nuclei.}
\label{bubble}
\end{center}
\end{figure}

Normalized spectroscopic factors, $C^2 S^{exp}_{norm}$, and proton
occupancies of the orbits are derived from the experimental cross
sections, as described in Section 'Methods'. The summed
spectroscopic factors for the first three $\ell$=2 states is 5.5
$\pm$ 1.0, consistent, within uncertainties, with a full occupancy
($2J+1$=6) of the $1d_{5/2}$ orbit. The weak population of the
3926 keV state, with a spectroscopic factor of 0.11(3), is tentatively
attributed to an $\ell$=2 proton knockout from the $1d_{3/2}$ orbit.
A proton occupancy of the $2s_{1/2}$ orbital of 0.17(3) is deduced
from the two observed $\ell$=0 excited
states at 2101 and 3704 keV. Assuming, though very unlikely, that the
four (unassigned) weakly populated states at 2666, 2696, 2816 and 3193 keV
were also $\ell$=0, this occupancy would be increased by only 0.07 (3).
This very small proton occupancy of the $2s_{1/2}$ orbital, 10$\%$
of that in $^{36}$S, results in a large depletion of the central density
of {\it protons}  in $^{34}$Si. 

On the other hand, the {\it neutron} $2s_{1/2}$
orbit is essentially fully occupied in $^{34}$Si, with a summed spectroscopic factor
value of 2.0(3) being deduced from the corresponding
neutron removal reaction from $^{34}$Si \cite{Enders}. Thus, $^{34}$Si
exhibits a large proton-to-neutron density asymmetry that, to our
knowledge, has not been revealed in any other nucleus. It is favored
because $^{34}$Si can be viewed as a doubly-magic nucleus in which
mixing between normally occupied and valence orbits is very limited
\cite{Enders,Burg14}. The high energy of its first excited state (3.3 MeV), its low reduced transition probability $B(E2; 0^+ \rightarrow 2^+)$ \cite{Ibbo} and the small electric monopole strength $\rho (E0; 0^+_1 \rightarrow 0^+_2$) \cite{Rota} complete this picture of double magicity. \color{black} Fig. \ref{bubble} visualizes the changed proton densities and almost unchanged neutron densities between $^{34}$Si and $^{36}$S from Relativistic Hartree-Fock Bogoliubov (RHFB) calculations that use the PKO2 energy density functional \cite{ebr11} and which predict very similar proton and neutron occupancies to those deduced here.  It should however be noted that mean field calculations do not all predict similar neutron and proton density profiles in $^{34}$Si. Indeed they are very sensitive to the size of the proton and neutron gaps derived from the choice of functionals, as well as to the treatment of pairing and quadrupole correlations that act to reduce the central density depletion, as found in Ref. \cite{Yao}. Indeed, this model-dependence of the predictions of the existence of a central depletion was a major motivation to perform the present experiment. 

With this differential two-fluid behavior, $^{34}$Si offers unique
possibilities to test the density and proton-to-neutron (isospin)
dependence of the nuclear spin orbit (SO) potential -- which
generates most of the shell gaps that stabilize magic nuclei in
the chart of nuclides \cite{SorlinPS, PPNP}. In most theoretical models, the
SO potential can be expressed in terms of the derivative of the
proton and neutron densities, with coefficients that differ
 by a factor of as much as two between various relativistic or non-relativistic
approaches (see e.g. the discussions in \cite{Sharma94,Sharma95,Lala98,Holt11,Bender,PPNP}).
These as yet unknown density and isospin dependences of the
SO interaction strongly impact (i) the evolution of the spin-orbit
interaction and magic numbers as one approaches the drip lines
\cite{Sharma94,Lala98}, where the surface diffuseness is increased
and consequently the SO interaction is expected to be reduced. This influences
the binding energies, the lifetimes, and nuclear capture rates of the nuclei
close to the neutron drip line that are involved in the synthesis of elements in
the Universe beyond Fe through the rapid neutron capture process.  This also
impacts (ii) the location of a possible island of stability for superheavy
nuclei \cite{Bender} that differ strongly depending on the theoretical models
used, and (iii) the puzzling discontinuity in the isotope
shifts observed for the Pb isotopes \cite{Sharma95,Rein95}, a phenomenon that seems to be accounted for only by a certain category of models.
These aspects of the SO force have not previously been accessible to experimental scrutiny as, in the vast majority of nuclei, the saturation of the nuclear forces implies (a) a near-constant central density for protons and neutrons, and (b) an almost universal surface diffuseness. The result is a SO force peaked at the proton and neutron surfaces having a similar strength for all models.
 The central proton density depletion in $^{34}$Si drives an additional
(interior) component of the SO force, with a 
sign opposite to that at the surface. Therefore, low-$\ell$ nucleons,
that can probe the interior of the bubble, should encounter a much
weaker overall SO force (e.g. \cite{Todd04,Bender}) and display a significantly
reduced SO splitting. This prediction is in line with the observed reduction of the
neutron $2p_{3/2} - 2p_{1/2}$ splitting in $^{35}$Si \cite{Burg14}, 
when compared to neighbouring $N=21$ isotones. Such a sudden change
by a factor of {\it two} in amplitude is unique on the chart of nuclides and
seems clearly connected to the change in central nuclear density observed here.   
Moreover, having different proton and neutron central densities, $^{34}$Si can be used to
constrain the isospin dependence of the SO interaction in an unprecedented 
manner, for example, by identifying models that predict the correct amplitude of
the SO reduction.

Finally, atomic nuclei are usually highly incompressible, the corresponding monopole modes involving very high excitation energies ~\cite{Hara,Young}. Exhibiting a central density that is significantly lower than the saturation density,  $^{34}$Si may present new (soft) compression modes at low energy with the potential to shed light on the recently observed fragmentation of the giant monopole at low energy in the neutron-rich $^{68}$Ni nucleus \cite{Vande14}. This information would in turn be useful for testing different models of the nuclear equation of state at a density below the saturation density, important for instance in the modeling of the neutron star crust.
\\

{\small
{\bf METHODS} \\
The eikonal model and choice of parameters used to calculate the
proton removal single-particle cross sections, $\sigma_\alpha^{sp}$,
and the parallel momentum, $p_{\parallel}$, distributions of the residues
are detailed in Ref. \cite{Gade08}. The shapes of the high momentum
parts of these distributions are used in the comparisons with experiment
in Fig. \ref{33Al-ls} as more dissipative collisions, treated only
approximately in the eikonal model, affect measured distributions
at the lower momenta \cite{Stroberg14}. In comparing to experiment,
the theoretical $p_{\parallel}$ distributions are convoluted with (i)
the momentum dispersion of the secondary beam, (ii) the beam straggling
in the target, and (iii) the momentum broadening due the reaction's
position within in the target.}

{\small The experimental partial cross sections, $b_f^{KO} \sigma_{inc,KO}^{exp}$,
correspond to the removal of a proton with quantum numbers $n\ell_J$ from
the ground state of $^{34}$Si. Here $\sigma_{inc,KO}^{exp}$ is the experimental
inclusive removal cross section, that amounts to 27.7(1.0) mb, and $b_f^{KO}$
is the experimental branching ratio (in $\%$) for populating final state $f$.
Following  Ref. \cite{36S}, the normalized knockout spectroscopic
factors are expressed as:
\begin{equation}
C^2 S^{exp}_{norm} = \frac{b_f^{KO} \sigma_{inc,KO}^{exp}}{R_s \sigma_f^{sp}},
\end{equation} \label{eq1}
where $\sigma_f^{sp} $ is the theoretical single-particle knockout cross section
\cite{Gade08}. $R_s$ accounts for the systematic quenching of measured
nucleon knockout cross sections when compared to those calculated when combining
these eikonal model $\sigma_f^{sp}$ with shell model spectroscopy and $C^2S$
\cite{Tostevin14}. With this normalization, the $C^2 S^{exp}_{norm}$ sum rule
(or orbit occupancies) to states in $^{33}$Al are normalized to the maximal
occupancy of a given sub-shell, that is $2J+1$. 

We are aware that short-range correlations~\cite{Pand97} and coupling to collective degrees of freedom \cite{Barb09} usually complicate the
determination of spectroscopic factors (or their related shell occupancies and
vacancies), which are not directly observable
\cite{Dick04,Furn02,Dugu15}. Moreover, present reaction models use effective
potentials that do not capture the full microscopic complexity of the nucleus
and often induce uncertainties in the deduced results. However, under the
reasonable assumption that these effects are similar between neighboring nuclei,
here between the closed-shell nuclei $^{36}$S and $^{34}$Si, the consideration
of a differential evolution of spectroscopic strengths and occupancies is
sensible, a view supported by tests of sum rules for occupancy and vacancy of
orbitals derived from experimental cross sections \cite{Schif12}. Error bars on
the occupancy values quoted in the main document include statistical and
systematical errors, the latter being derived from Ref. \cite{Tostevin14}.

We note that the measured inclusive cross section for the removal of an $\ell=0$ proton is about 12 times larger in $^{36}$S than in $^{34}$Si. This directly measured cross-section ratio, attributed to the almost complete depletion of the $2 s_{1/2}$ proton orbit between the two nuclei, has a value that is very similar to the ratio of occupancies derived from the reaction model calculations presented. 

It is hoped that the present study will stimulate new developments in the modeling of nuclear reactions and their application to nuclear spectroscopy and further motivate the construction of a high-luminosity electron - radioactive nuclei collider facility that would enable a more direct experimental determination of the proton density distribution in $^{34}$Si.
\\ 

{\scriptsize
{\bf Data availability}: Raw data were obtained at the Coupled Cyclotron Facility at the National Superconducting Cyclotron Laboratory, Michigan State University, USA. All other derived data used to support the findings of this study are available from the authors upon request and a thorough explanation of the analysis method can be found in Ref.\cite{these}}.\\

{\scriptsize
{\bf Acknowledgments} This work is supported by the National Science Foundation (NSF) under Grant Nos. PHY-1102511 and PHY-1306297, the OTKA Contract No. K100835, as well as by the Institut Universitaire de France. GRETINA was funded by the US DOE - Office of Science. Operation of the array at NSCL is supported by NSF under Cooperative Agreement PHY-1102511 (NSCL) and DOE under grant DE-AC02-05CH11231 (LBNL). J.A.T acknowledges support of the Science and Technology Facility Council (UK) grant ST/L005743. \\

{\bf Author Contributions} A.M performed the off-line data analysis,  A.Lem, D.W, K.W performed on-line data analysis and checked the integrity of data taking. K.W, A. Lem performed GEANT4 simulations and wrote parts of the offline sorting code. A.G and J.T performed reaction theory calculations. D.B operated the S800 spectrometer. D.W and F.R were responsible for the setting up, calibration and operation of the Gretina array. H.I,K.W helped to setup the Gretina array. The manuscript was prepared by O.S, A.M, J.T,  A.G, A.Lem and E.K. J-P.E performed relativistic mean field calculations. Z.D and D.S contributed to the off-line data analysis of the $\gamma$-ray spectra and C.B, R.B, E.K, A.Lem, A.Lep, H.I, T.R, M.S, M.V, R.S checked data accumulation on-line. O.S. proposed the experiment and supervised the analysis.\\

{\bf Author Information} The authors declare no competing financial interests. Readers are welcome to comment the online version of the paper. Correspondence and requests for material should be addressed to O.S. (sorlin@ganil.fr).

}

\end{document}